\documentclass[twocolumn,floatfix]{aastex63}

\usepackage{amssymb}
\usepackage{amsmath}
\usepackage{hyperref}
\usepackage{utfsym}

\newcommand*\unicorn{\vcenter{\hbox{\includegraphics[width=1em]{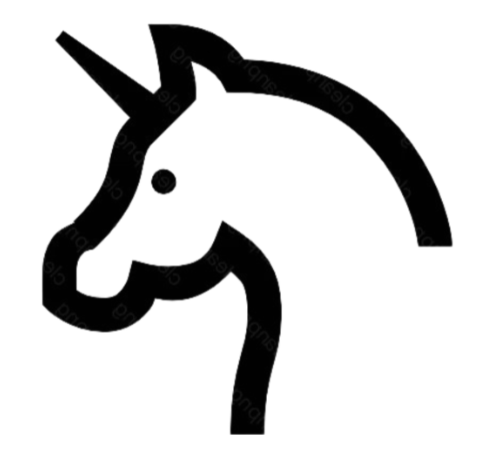}}}}

\usepackage{graphicx}

\begin{document}

\title{Dark Energy Constraints and Joint Cosmological Inference from Mutually Inconsistent Observations}

\correspondingauthor{Charles Steinhardt}
\email{steinhardt@nbi.ku.dk}

\author[0000-0003-3780-6801]{Charles L. Steinhardt}
\affiliation{Department of Physics and Astronomy, University of Missouri, 701 S. College Ave., Columbia, MO 65203}
\affiliation{Dark Cosmology Centre, Niels Bohr Institute, University of Copenhagen, Jagtvej 155A, DK-2100 Copenhagen \O}

\author[0009-0009-5025-8134]{Preston Phillips}
\affiliation{Department of Physics, Math, and Astronomy, University of Texas, 2515 Speedway, Austin, TX 78712}
\affiliation{Dark Cosmology Centre, Niels Bohr Institute, University of Copenhagen, Jagtvej 155A, DK-2100 Copenhagen \O}

\author{Radoslaw Wojtak}
\affiliation{Dark Cosmology Centre, Niels Bohr Institute, University of Copenhagen, Jagtvej 155A, DK-2100 Copenhagen \O}


\begin{abstract}
Recent results regarding dark energy are mutually inconsistent under the $\Lambda$CDM cosmological model, hinting at the possibility of undiscovered physics. However, the currently accepted cosmological parameters come from a joint inference between observational data sets, a process that is formally invalid for inconsistent observations. We will show that many problems can arise when using joint inference on disagreeing observations such as significantly overestimated margins of error, high dependencies on priors, and sensitivities to boundary constraints. Because we do not know if these inconsistencies arise due to errors in observation, poor statistical techniques, or an improper model, it is difficult to fix these problems. We will discuss each scenario in which the analysis method breaks and explore an alternative resampling technique, developing methods that may make determining the sources of tensions in cosmological parameters easier.
\end{abstract}


\section{Introduction}
\label{sec:intro}

Over the past few decades, a concordance cosmological model ($\Lambda$CDM) has emerged in which the Universe is composed primarily of a cosmological constant ($\Lambda$) associated with dark energy, cold dark matter, and baryons.  The current best-fit $\Lambda$CDM model is derived by combining observations using disparate techniques and ranging from the local Universe to the time of primordial nucleosynthesis.  Each of these individual measurements is typically described in terms of the constraints placed on the parameters characterizing the expansion history and growth of structure in a $\Lambda$CDM model: the Hubble constant $H_0$ and density parameters $\Omega_\Lambda$ and $\Omega_m$ for dark energy and matter, respectively.  Then, a joint Bayesian inference determines the best-fit parameters.

Although the result is described as constraints on $\Lambda$CDM parameters, the individual observations instead primarily constrain combinations of other parameters.  For example, \emph{Planck} measurements of the cosmic microwave background (CMB) \citep{Planck2020} can be approximated as a covariance matrix between the baryon density $\Omega_b h^2$ and two shift parameters, $\mathcal{R}$ and $\theta_*$ \citep{Efstathiou1999}.  The parameters most directly probed by local standard candles, baryon acoustic oscillations (BAO), and CMB experiments comprise disjoint sets.  Thus, a cosmological model must be assumed in order to convert them to a common basis in order to produce a joint inference.  

One symptom of an incorrect cosmological model would be that measurements using different techniques appear to be mutually inconsistent.  Before the introduction of dark energy in the late 1990s, measurements indeed appeared to be mutually inconsistent \citep{Ostriker1995}.  The current $\Lambda$CDM model reconciled all of the existing measurements \citep{Riess1998,Perlmutter1999}, which is why it has been termed a concordance cosmological model.  

Over the past few years, improved precision in cosmological observations has again produced possible inconsistencies between cosmological observations using different techniques.  There is now a `Hubble tension' between measurements of $H_0$ from local and high-redshift observations \citep{Riess2022}.  Recent BAO observations also hint at the possibility that a model in which dark energy has a time-varying equation of state $w(z)$, where $z$ is the redshift, may be statistically preferred over $\Lambda$CDM, for which $w = -1$ \citep{DESIY1}.  These results, like previous analyses, use established techniques for joint inference.

If it can be robustly established that the dark energy equation of state is not constant, this would be the first meaningful property of dark energy discovered since its introduction.  Our predictions for the future expansion trajectory and even the ultimate fate of the Universe would depend upon the result.  Given the importance of the result, it is particularly essential that any conclusions drawn about dark energy be robustly established using techniques with a rigorous mathematical foundation and resilient to changes in priors and model assumptions.  If this is not possible with current techniques, one must develop improved techniques in order to properly explore ideas for reconciling current observations.

The most common technique currently used for joint inference begins by setting priors on cosmological parameters.  Each observational constraint is described as a likelihood.  The posterior probability distribution is then calculated by combining all of these likelihoods with the prior, often via a sampling algorithm since direct integration of probability density in high-dimensional spaces is computationally intensive.  

However, this relies on the fundamental assumption that the measurements being combined are mutually consistent.  If they are inconsistent, as hinted at by recent results, then the posterior probability can produce a wrong inference.  In the case of dark energy, an incorrect inference could mean the difference between predicting that the Universe will expand forever and predicting that it will eventually turn around and begin contracting.  This work demonstrates several potential dangers of joint inference from inconsistent cosmological observations and discusses possible solutions.

The testbed used for exploring cosmological inference in this work is described in \S~\ref{sec:methods}.  In \S~\ref{sec:tension}, it is shown that incorrectly combining inconsistent constraints such as in current observations can lead to an underestimate of statistical uncertainties.  Several possible resolutions are then considered. 
 If the observations are instead considered in isolation, it is shown that the result will have a strong dependence upon the initial priors and bounds selected (\S~\ref{sec:priors}).  The difficulties arising from approximating a complex combination of statistical and systematic uncertainties with a single covariance matrix are described in \S~\ref{sec:systematic}.  In \S~\ref{sec:model}, the possibility of resolving these issues in a similar manner to the late 1990s, by searching for an improved cosmological model that can again produce concordance, is explored.  The implications of these results for current and upcoming observations are discussed in \S~\ref{sec:discussion}.

\section{Current Techniques for Bayesian Joint Inference}
\label{sec:methods}

Currently, cosmological parameters are determined by combining multiple observations at a wide range of redshifts.  This sort of joint inference is common because each individual dataset is only sensitive to some components of the $\Lambda$CDM model.  For example, because the $\Lambda$CDM dark energy density is negligible at the time of the CMB, CMB observations alone do not produce strong constraints on $\Omega_\Lambda$ in the absence of additional assumptions such as flatness.

Each observational result is treated as an independent measurement constraining different combinations of cosmological parameters.  In order to use them to produce a joint constraint, each is described as a likelihood distribution.  A prior function $p(\theta)$ is chosen, and Bayesian joint inference can then be performed to determine the posterior probability distribution. It should be noted that this is only valid if the likelihoods are independent measurements of parameters belonging to the same model.  For mutually inconsistent measurements, this joint inference will be invalid.

Combining likelihoods is complex in several ways.  First, the data place constraints on distinct properties which are related by a common model.  For example, Planck measures the baryon density $\Omega_b h^2$, where $h = H_0/{100 \textrm{km/s/Mpc}}$, more precisely than $H_0$, which is better constrained from local standard candles.  Thus, the model which is being tested is a fundamental assumption in the analysis, and the posterior distribution obtained using one model cannot be used to constrain parameters under a different model.

Further, because a computationally-expensive model must be run each time a new set of parameters is sampled, it is necessary to produce the posterior distribution from a limited number of samples.  For this reason, a randomized Monte Carlo Markov chain (MCMC; \citep{Metropolis1953})  is used to explore the posterior probability space. 

In the typical cosmological usage, MCMC combined with the \texttt{emcee} sampler \citep{ForemanMackey2013} uses walkers to create an approximate posterior probability distribution for the joint likelihoods. On initialization, each walker begins at a point in the probability space. Then, a cosmological theory code, such as \texttt{camb} \citep{Lewis2002} or \texttt{classy} (CLASS; \citealt{Blas2011}), is combined with the given likelihoods and priors to calculate the posterior at that point. Each walker moves a random direction in the probability space (expressed as a vector). The posterior is calculated at this new point. If the new point is more probable, the walker takes the step. If not, the walker rejects the step with a certain probability, as with simulated annealing algorithms \citep{Metropolis1953}. This is to prevent the walker from getting stuck in local minima/maxima.  

At regular intervals, the distribution of walkers is tabulated and compared with the previous tabulated distribution.  Once the two are sufficiently similar to within some tolerance, the process is deemed to have converged.  The final distribution of walkers at convergence is used as an estimate of the posterior probability density function.

The process results in an approximation of the posterior distribution centered around the points with the highest probability density. The triangle graphs often shown as a result of Bayesian analysis show two-dimensional slices of the probability distribution for each pair of parameters of interest. The final values for each parameter are calculated by taking the mean of the posterior probability distribution with uncertainty expressed as the upper and lower bounds of a confidence interval.

The examples in this paper use the \texttt{cobaya} \citep{Torrado2021} framework with CosmoMC MCMC sampler \citep{Lewis2002} to perform cosmological inference, further relying on the \texttt{camb} \citep{Lewis1999} code for cosmological model calculations.  However, the issues discussed here are independent of the MCMC implementation used.

Finally, each set of observations can produce a complex likelihood that is difficult to probe.  In order to make the problem tractable, these are typically transformed into a covariance matrix between a collection of cosmological parameters.  This approximation assumes that all relevant uncertainties can indeed be described as a covariance matrix, which is not true for systematic errors.  

Of the various assumptions made by this standard technique, two are particularly difficult to validate.  First, each constraint relies on complex analytical techniques leading to systematic errors which should not be ignored and cannot be described in terms of a covariance matrix.  Second, current cosmological measurements appear to be mutually inconsistent, which, even if it were possible to handle the full probability distribution including systematic uncertainties properly, would invalidate the Bayesian joint inference.  In \S~\ref{sec:tension}, each of these are explored in more detail.

\subsection{Cosmological Datasets Used}

In order to provide concrete examples, this work will primarily focus on the combination of three current observational constraints: supernova observations from Pantheon+ \citep{Scolnic2022}, BAO observations from the DESI Year 1 sample \citep{DESIY1}, and CMB observations from Planck \citep{Planck2020}.  The covariance matrices for these constraints are implemented as part of the \texttt{cobaya} framework. 

\section{Joint Inference from Conflicting Measurements}
\label{sec:tension}

As described above, the joint inference used assumes that all differences between measurements are due to imprecision.  That is, with larger samples, the various techniques for constraining cosmological parameters would converge to the a common set of $\Lambda$CDM parameters.  However, currently this does not appear to be the situation.  Rather, the tension between, e.g., $H_0$ as measured from type Ia supernovae calibrated with direct geometric distance measurement via Cepheid observations and from early-Universe observations \citep{Riess2022} is very unlikely to be a particularly improbable $\sim 5\sigma$ deviation that will disappear solely with improved sample size.  Instead, it is likely that the two measurements are truly inconsistent under $\Lambda$CDM.  If so, the differences between they are not solely due to measurement imprecision, so it is no longer valid to apply both likelihoods as updates to a single prior.

\begin{figure}[h]
    \centering
    \includegraphics[width=0.4\textwidth]{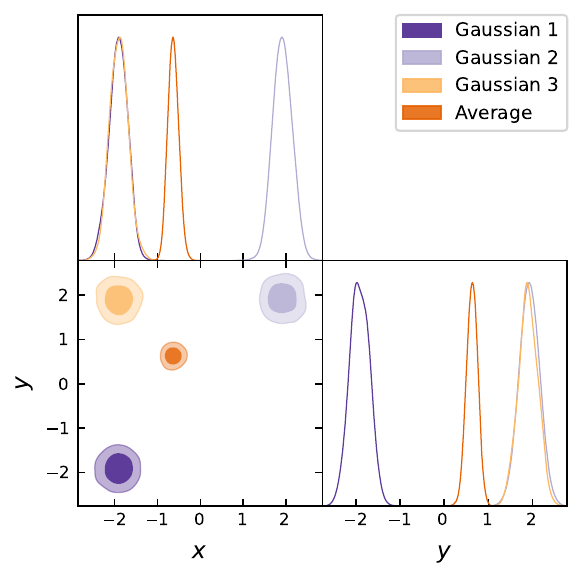}
    \caption{The joint inference from three inconsistent Gaussians (Gaussian 1, 2, and 3) produces a new Gaussian distribution centered at the average.  The small inferred uncertainty comes from the assumption that all deviations from the true parameters are due to random measurement imprecision.  However, in practice this is unlikely to be the case for discrepant results, in which case the uncertainty will be underestimated, and the true parameters will very often lie much further from the average.}
    \label{fig:gaussian} 
\end{figure}

To demonstrate the effect, consider the joint inference from three inconsistent measurements with identical, normally-distributed uncertainties, as illustrated in Fig. \ref{fig:gaussian}.  The central values for inferred parameters, as might be expected, lie at the average of the three measurements.  However, the inferred uncertainty is quite small, and each individual measurement is predicted to be a significant outlier.  

This small inferred uncertainty is due to the assumption that all deviations from the true parameters are due to random measurement imprecision.  As unlikely as it is to have randomly measured three, e.g., $5\sigma$ outliers, this is still far more probable than having one $2 \sigma$ outlier and two $8\sigma$ outliers.  Thus, the posterior probability density function is narrowly centered around the average value.  

\begin{figure}[h]
    \centering
    \includegraphics[width=0.4\textwidth]{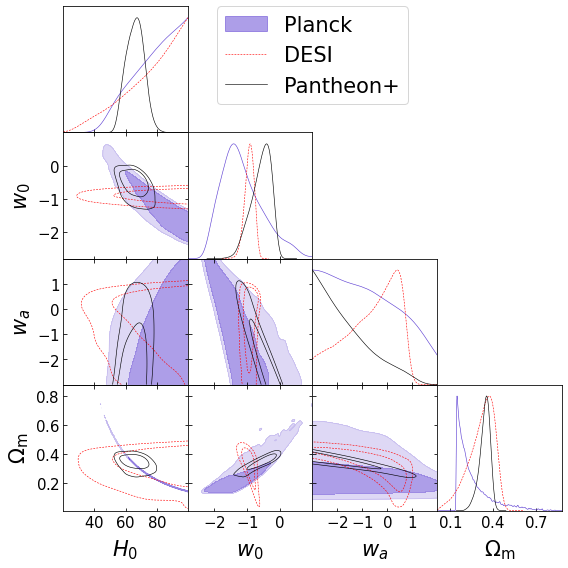}
    \caption{Each independent inference obtained from Planck CMB, DESI BAO, and Pantheon+ Supernovae under the $w_0w_a$ cosmological model. Although all three are consistent with the possibility of a non-$\Lambda$CDM cosmology, they prefer different $w_0w_a$ parameters, and deviate from each other more than should be expected from statistical uncertainties alone.  One possible explanation could be underestimated uncertainties.}
    \label{fig:w0wa_independent} 
\end{figure}

A similar, although less severe, problem exists due to the discrepancies between current cosmological measurements.  For example, a joint analysis of Pantheon+, DESI, and Planck was used by \citet{DESIY1} to infer the possibility of a variable dark energy equation of state, with $w(a) = w_0 + (1-a)w_a$.  However, for the $w_0w_a$ model, the individual measurements are inconsistent, preferring different  parameters (Fig. \ref{fig:w0wa_independent}).  Thus, the uncertainties on best-fit $w_0w_a$ parameters are almost certainly underestimated.  In the following section, several possible remedies are considered.

\section{Potential Solutions and Primary Difficulties}
\label{sec:solutions}

Joint inference from conflicting measurements is thus not possible when they are all treated as independent measurements of the same cosmology.  The proper resolution depends upon the root cause of the measured inconsistency in constraints.  Here, three possible explanations are considered:
\begin{itemize}
    \item One or more measurement techniques could be fundamentally flawed, likely due to assumptions necessary to perform the analysis.  In that case, it would be necessary to discard those constraints entirely and constrain cosmological parameters solely from the remaining measurements.
    \item Some techniques could have significant systematic uncertainties which have not been properly accounted for.  If so, a proper inclusion of those systematic uncertainties could again produce concordance, in which case current techniques for joint inference would produce a valid result.
    \item The inconsistency between measurements when assuming $\Lambda$CDM could indicate that a different cosmological model is required.  The proper resolution would then be to search for a new model which produces concordance, much as was done during the development of $\Lambda$CDM.
\end{itemize} 

Each of these potential resolutions is discussed in the remainder of this section.  All three have potential complications which must be addressed in order to produce a robust result.

\subsection{Choice of Priors and Bounds for Single Measurements}
\label{sec:priors}

Perhaps the most straightforward explanation for conflicting measurements is that some measurements are flawed.  If so, the best solution would naturally be to only draw conclusions from the unflawed measurements.  Unfortunately, this turns out to be more complex than it might first appear.  

First, one must determine which measurements to discard.  If, for example, most measurements produced a consistent answer and one measurement disagreed, it would be logical to guess that the discrepant measurement should be the one to discard.  The problem, though, is that this is poor statistical practice and potentially dangerous for the same reason that clipping outliers is often flawed.  Removing data solely on the basis that they are discrepant under $\Lambda$CDM would make it impossible to discover true discrepancies that might indicate the need for a different cosmological model.  Thus, one should only remove data on the basis of identifiable flaws, systematic uncertainties, or other biases.  This can be complex since nearly every astronomical dataset relies on strong assumptions that are difficult to verify in isolation. 

Further, even pairwise, any two out of supernova observations, BAO observations, and CMB observations currently exhibit tension under $\Lambda$CDM.  Thus, the most conservative approach would be to simply consider each dataset in isolation.  

Results relying solely on a single dataset are typically presented along with a joint inference combining multiple datasets.  However, $\Lambda$CDM parameters cannot be well constrained from any single dataset due to the time evolution of the composition of the Universe.  For example, at a redshift $ \sim 1100$, around the time of CMB emission, the dark energy to dark matter ratio will be $< 10^{-9}$ of the current ratio.  Thus, CMB observations provide negligible information on dark energy and its properties.  For alternative models in which dark energy could have a varying equation of state, CMB constraints are also asymmetric since there will be effectively no constraints as $w \rightarrow -\infty$, but a large $w$ would lead to a dark energy density high enough to have observable consequences.

One might expect that parameters which are poorly constrained by the data should simply produce a broad posterior distribution.  However, for a parameter which is entirely unconstrained, the true posterior distribution will be unbounded.  Thus, the probability density should approach zero at every point\footnote{More rigorously, the posterior distribution does not fit the definition of probability distribution, much in the same way that one cannot place a flat, uniform prior on the full set of real numbers.  In practice, for fixed, non-zero tolerance, given unlimited computation time MCMC will eventually terminate in a well-defined, bounded, and incorrect probability density.}.  Further, it would take infinite computational time to reach this solution. 

As a result, in addition to choosing a prior probability distribution, MCMC walkers are limited to a bounded parameter space.  Ideally, these bounds are chosen so that anything outside of the space is so physically unreasonable that it should not be worth considering as a possible cosmology.  For example, the default \texttt{cobaya} bounds on $w_a$ in $w_0w_a$ cosmology are $-3 \leq w_a \leq 1$.  The upper bound is chosen because assuming $w_0 \approx -1$, at $w_a = 1$, dark energy would become degenerate with dark matter at very high redshift, such as for CMB observations.  The lower bound is chosen because, for $w_a \lesssim -3$, the ages of stellar populations would be larger than the age of the Universe.

This latter argument, however, has a poor cosmological history.  Stars which appear to be older than the age of the Universe have previously been reported  \citep{Bolte1995,Bond2013}.  However, they typically instead can be explained by uncertainties in the difficult measurement of stellar ages.

Further, one could easily argue for a more restrictive bound.  Models with $w < -1$ would require exotic physics, including a negative kinetic energy \citep{Caldwell2002}.  This is most commonly done by assuming that the kinetic energy for a phantom dark energy field $\phi$ is proportional to $-\frac{1}{2}\dot{\phi}^2$, rather than the standard $\frac{1}{2}\dot{\phi}^2$.  However, this term will then be negative definite instead of positive definite.  Developing a model which can cross $w = -1$ requires something even more complex, such as the introduction of a Lagrangian term proportional to $\Box^2$, where $\Box = \partial^\mu\partial_\mu$ \citep{Li2005}.  Thus, perhaps a bound at $w_a \geq - 1 - w_0$ would be a reasonable choice.  Alternatively, one could also argue for a less restrictive bound than $w_a \geq -3$ in order to make fewer assumptions about plausible theoretical models.  

The problem is that the non-local observational constraints do provide an upper bound on $w_a$, but not a lower bound.  Thus, if the walkers are not bounded, they will continue to reach progressively lower values of $w_a$, and the posterior probability distribution calculated from those walkers will tend towards $w_a \rightarrow -\infty$.  For example, DES, allowing $w_a$ to be a low as -15, reports a best-fit $(\Omega_m, w_0, w_a) = (0.495^{+0.033}_{-0.043},-0.36^{+0.36}_{-0.30},-8.8^{+3.7}_{-4.5})$ \citep{Abbott2024}.  However, this is strongly dependent upon their choice of priors and bounds, as well as the tolerance used to decide when the MCMC chain has converged.

\begin{figure*}[h]
    \centering
    \includegraphics[width=0.9\textwidth]{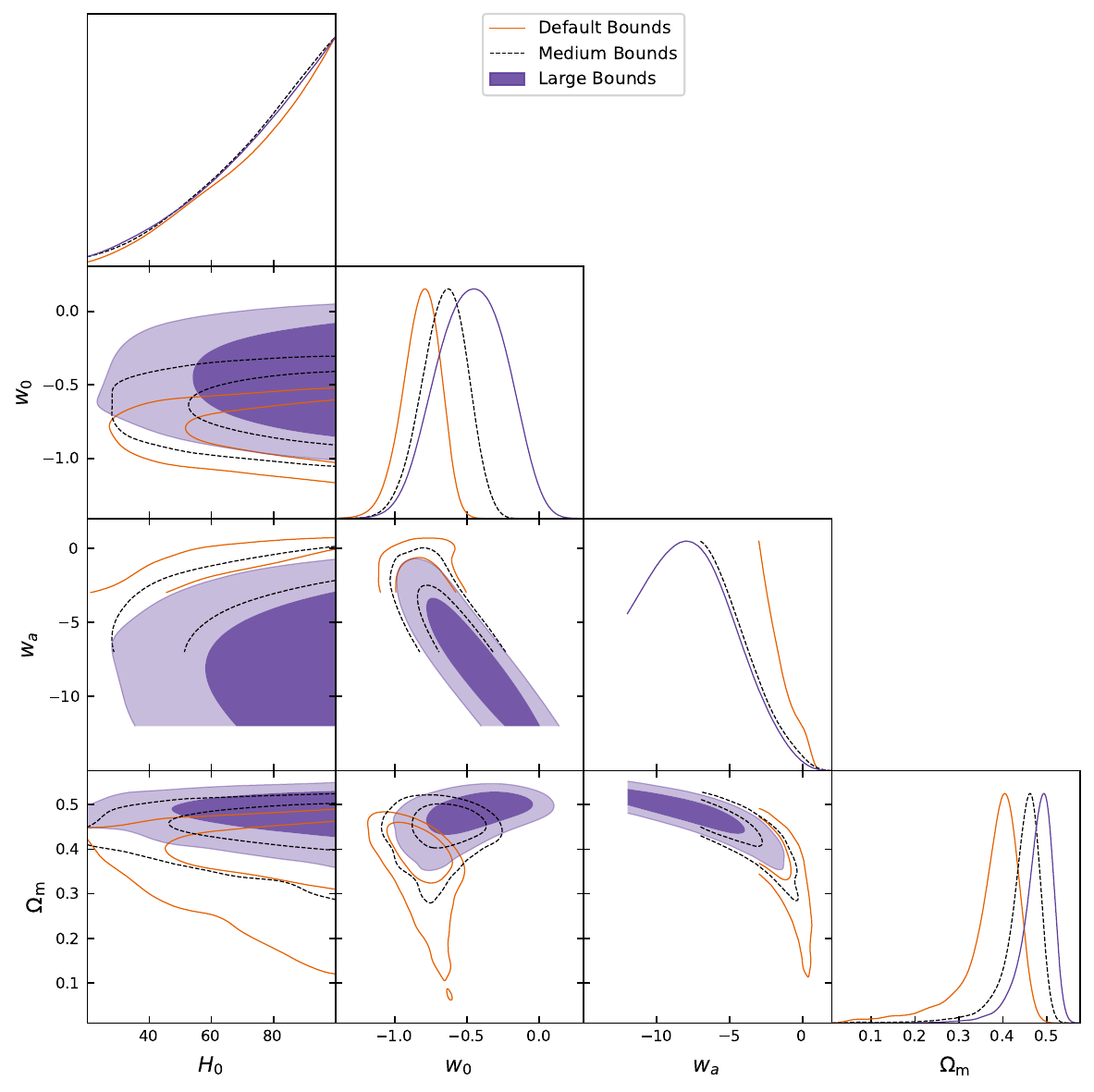}
    \caption{The inferred posterior probability distributions for the DES Year 5 Supernova dataset depend strongly on the bounds chosen for $w_a$. The default bounds in cobaya correspond to the first row of Table \ref{tab:bounds}, with the medium bounds corresponding to those from the second row, and large bounds from the third.  The resulting posterior probability distributions are highly sensitive to the choice of bound, even though the bulk of the distribution does not lie near the bound.}
    \label{fig:bounds} 
\end{figure*}

When re-fitting the Dark Energy Survey (DES) Year 5 supernova data with an allowed range of $[-3,2]$, with 95\% confidence $-3 < w_a < 0.16$.  However, if the allowed range is increased to $[-12,2]$, with the same confidence $-12 < w_a < -2.15$ (Table \ref{tab:bounds}; Fig. \ref{fig:bounds}).  Thus, depending upon whether the stellar age-based bound at $w_a > -3$ is assumed, one might either conclude that DES is consistent with $w_a = 0$ and a constant dark energy equation of state or that there is $>2\sigma$ evidence for a variable equation of state.  Even a result of such fundamental physical importance is not robust against a change in priors.

\begin{table*}[hbt!]
    \centering
    \begin{tabular}{cccccc}
        $w_0$ Bounds & $w_a$ Bounds & $H_0$ & $\Omega_m$ & $w_0$ & $w_a$ \\
        \hline
        [-3,1] & [-3,2] & $76.0^{24.0*}_{-7.0}$ & $0.374^{+0.071}_{-0.024}$ & $-0.819^{+0.145}_{-0.112}$ & $-1.74^{+0.34}_{-1.26*}$ \\
        $[-7,2]$ & $[-7,2]$ & $75.8^{24.2*}_{-7.0}$ & $0.441^{+0.049}_{-0.017}$ & $-0.662^{+0.175}_{-0.164}$ & $-4.59^{+0.61}_{-2.41*}$ \\
        $[-12,3]$ & $[-12,2]$ & $75.2^{24.8*}_{-6.9}$ & $0.473^{+0.046}_{-0.017}$ & $-0.471^{+0.244}_{-0.243}$ & $-7.29^{+1.98}_{-3.94}$ \\
    \end{tabular}
    \caption{  Best-fit parameters for a $w_0w_a$ model \citep{Chevallier2001,Linder2003,dePutter2008} from DES Type Ia SNe data using different cobaya bounds.  Apart from the bounds, the data and other settings are identical.  The choice of bounds not only changes the central values for cosmological parameters, but even determines whether the resulting fit is consistent with $\Lambda$CDM and whether it is consistent with, e.g., CMB constraints on $\Omega_m$.  An asterisk indicates that the 68\% contour is not measured, but instead reaches a prior bound, either in $H_0$ at 100 km/s/Mpc or the indicated bound in $w_a$.  DES does not place strong upper bounds on $H_0$.}
\label{tab:bounds}    
\end{table*}

In principle, one might make an argument for a particular choice of bound, such as the stellar age argument used to bound $w_a > -3$ in the default \texttt{cobaya} parameters.  However, this leads to a series of additional problems that are likely best considered as philosophical rather than astrophysical, discussed in \S~\ref{sec:discussion}.   The most conservative approach is to consider that none of these bounds is indisputably or falsifiably correct, and therefore any conclusion which relies on a choice of bounds is not truly supported by the data.

\subsection{Systematic Uncertainties}
\label{sec:systematic}

In order to constrain the parameters in the dark energy model, all three data sets (CMB, BAO, and supernova) must be used to produce a joint inference.  The fundamental assumption made by current joint analysis techniques is that all differences between the true underlying cosmological parameters and measurements are solely due to measurement imprecision.  However, under $\Lambda$CDM and several related models, these data are not mutually consistent.  More precisely, given current measurements and reported uncertainties, it is highly unlikely that these three datasets are truly measuring the same $\Lambda$CDM parameters.

However, this conclusion relies on the reported uncertainties, and there is a long history of underestimated uncertainties in cosmological measurements \citep{Tully2023}.  If uncertainties have been underestimated, then it is possible that a single set of $\Lambda$CDM parameters could fit all existing observations, in which case $\Lambda$CDM would still be a concordance model.  In particular, current observational constraints are transformed into covariance matrices for use as a likelihood in joint inference.  However, a systematic uncertainty cannot be described as covariance.  Thus, systematic errors are not properly represented using current methods.  

For example, constraining the Hubble constant from supernova measurements requires calibration against other standard candles.  Pantheon+ is typically paired with SH0ES \citep{Riess2022} in order to produce a calibrated dataset.  However, Pantheon+SH0ES is inconsistent with early-Universe measurements.  As a result, the joint analysis produced by DESI instead relies on uncalibrated SNe, using Pantheon+ (or several alternative supernova datasets) without SH0ES calibration.  

This would be logical if one assumes that the Hubble tension is due to errors in SH0ES.  Although this is possible, it is by no means the only plausible explanation for the Hubble tension.  Moreover, removing SH0ES entirely is difficult to justify.  Even if SH0ES has significantly underestimated systematic uncertainties, this would not mean that SH0ES carries no information.  

Further, even without SH0ES, constaints from SNe, BAO, and CMB observations appear to be inconsistent under a single set of $\Lambda$CDM parameters, instead favoring a more complex model such as a varying dark energy equation of state \citep{DESIY1}.  Thus, supernova calibration cannot be solely responsible for the existing tension and discarding SH0ES does not provide a solution.  
A possible solution is to assume that the tension is due to underestimated uncertainties \emph{without specifying a specific set of observations as being primarily responsible}.  That is, one can return to the core assumptions behind Bayesian joint inference: (1) that $\Lambda$CDM is the correct cosmological model; and (2) that the only discrepancies between the true $\Lambda$CDM parameters and observations are due to measurement imprecision.  However, a nonparametric method is adopted in order to remove the assumption that the given covariance matrices fully represent that measurement imprecision.  It is important to note that because a specific model (whether $\Lambda$CDM or another) is assumed, this method can no longer be used as a test of that model. Rather, it can only be used to estimate the confidence intervals for parameters within that model assuming that it is correct.

Resampling is a standard non-parametric technique for dealing with unknown uncertainties. Although the use of likelihoods in this joint inference makes conventional resampling not possible, a variation of these methods may allow for the creation of more accurate uncertainties.

The analysis is performed by repeatedly drawing sets of $N$ out of the $N$ available likelihoods with replacement.  For example, a draw using Pantheon+, SH0ES, DESI, and Planck might select the DESI likelihood twice, Planck, and Pantheon+.  The joint inference is then performed, producing a set of MCMC chains.  This is repeated a large number of times, and the chains from each draw are combined to produce a single posterior distribution.  For a small $N$, as is available here, instead of randomized trials, a more exact posterior can be produced by calculating chains for all of the available combinations, weighted by the multiplicity with which that combination is chosen in a random draw.  

\begin{figure}[h]
    \centering
    \includegraphics[width=0.4\textwidth]{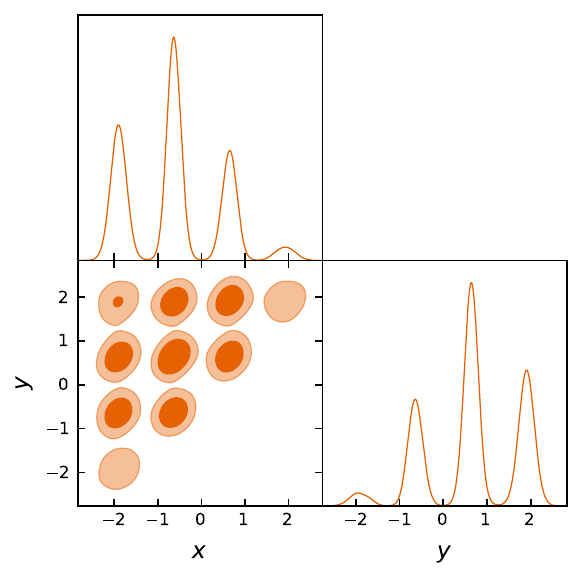}
    \caption{The same three Gaussian joint inference from Fig. \ref{fig:gaussian}, resampled over each possible weighting of the Gaussian distributions. This yields a much broader confidence interval than the original joint distribution, since it is assumed via resampling that the seemingly inconsistent original measurements must truly be consistent, and therefore it can be inferred that their stated uncertainties are underestimates.}
    \label{fig:resampling} 
\end{figure}

For mutually inconsistent observations, this will generally produce a larger posterior distribution than one would get using the given uncertainties (e.g., Fig. \ref{fig:resampling} for the Gaussian example from \S~\ref{sec:tension} and Fig. \ref{fig:gaussian}).  The opposite would be true in a situation where the reported uncertainties are overestimated.

\begin{figure}[h]
    \centering
    \includegraphics[width=0.4\textwidth]{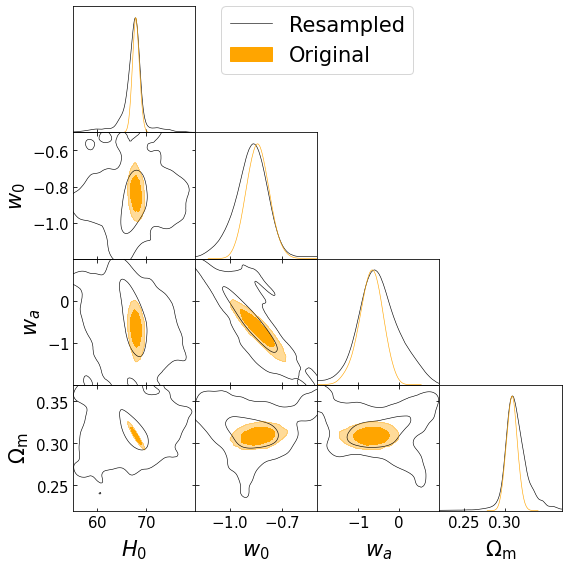}
    \caption{The $w_0w_a$ model resampled over the possible combinations of the three data sets.  The resulting posterior distribution is significantly broader than a joint inference relying on the stated uncertainties, and is consistent with $\Lambda$CDM.  Note that a handful of possible resampling draws were omitted because they would diverge, thus pulling the joint inference for some parameters towards (negative) infinity.}
    \label{fig:resampled_cosmology} 
\end{figure}

Applying this technique to Pantheon+SH0ES, DESI, and Planck produces the posterior distribution shown in Fig. \ref{fig:resampled_cosmology}.  In this technique, it is explicitly assumed that all of these observations consist of measurements of the same $\Lambda$CDM parameters, and that the apparent inconsistencies were due to underestimated uncertainties.  Thus, by construction a best-fit $\Lambda$CDM model with a posterior distribution broad enough to be consistent with all of the included observations is produced.  

However, this also means that if the Hubble tension or BAO observations truly indicate a non-$\Lambda$CDM cosmology, resampling will instead continue to expand the uncertainties until some $\Lambda$CDM model is reached.  It would not be possible to use this technique to falsify $\Lambda$CDM or any other model which is imposed.  

\subsection{Choice of Model}
\label{sec:model}

The final possibility to consider is that observations might truly be inconsistent under $\Lambda$CDM, indicating that a different cosmological model is needed.  This would be the most intriguing resolution, as it would likely require new fundamental physics.

Given that there are a limited number of observations which constrain cosmology at very different epochs, it will always be possible to produce a sufficiently complex model to bring everything into concordance.  Indeed, there should be an infinite family of such models.  The goal is therefore not merely to find a model that can fit all of the observations, but to find a physically motivated model with a small number of parameters which is still successful.

A natural candidate is that dark energy might not be cosmological constant, but rather have dynamics and a variable equation of state.  Since the deviation from $\Lambda$CDM found by DESI is concentrated in their lowest-redshift observations, a plausible candidate might be a thawing quintessence model (cf. \citep{Tsujikawa2013}).  Here, a scalar field $\phi$ is introduced, which in an expanding Universe will evolve as 
\begin{equation}
    \Ddot{\phi} + 3H(t)\dot{\phi} + V^\prime(\phi) = 0,
\end{equation}
where $H$ is the Hubble parameter and $\dot{x}$ and $x^\prime$ represent the derivatives with respect to time and $\phi$, respectively.  The dark energy equation of state will be 
\begin{equation}
    w_\phi = \frac{p}{\rho} = \frac{\frac{1}{2}\dot{\phi}^2 - V(\phi)}{\frac{1}{2}\dot{\phi}^2 + V(\phi)}
\end{equation}
in natural units.

At high redshift, dark energy is a negligible fraction of the energy density $\rho$, so unless $V^\prime \gg V$, $H \propto \rho^{1/2} \gg V^\prime(\phi).$  Thus, $\dot{\phi}$ is small and $w_\phi \approx 1.$  As the Universe expands and dark energy begins to dominate, $V^\prime(\phi)$ is no longer negligible and $\phi$ begins to evolve, leading to an increase in $w_\phi$.  The details of this evolution depend upon the potential $V(\phi)$, and for the right choice of potential, nearly any behavior can be produced.

Any potential $V(\phi)$ specifies a model which can then be used to produce a joint inference.  However, that inference will only be valid for the specific model which is tested.

In an attempt to consider a generic version, it is standard to introduce the approximation
\begin{equation}
w_\Lambda = w_0 + w_a (1-a),    
\end{equation}
where variation is approximated as linear in the scale factor $a$ \citep{Chevallier2001,Linder2003,dePutter2008}.  In such a model, the present value of $w$ is $w_0$ and the early-Universe value approaches $w_0 + w_a$.  

Although it might seem intuitive to use a linear approximation for an arbitrary function, the typical evolution of a thawing quintessence model is highly non-linear.  This leads to particularly counterintuitive results at high redshift, where dark energy should be a negligible fraction of the energy density and therefore the equation of state is not well constrained.  For the best-fit joint DESI, CMB, and Pantheon+ fit of $w_0 = -0.831 \pm 0.066, w_a = -0.73 ^{+0.32}_{-0.28}$ \citep{DESIY1}, the high-redshift dark energy equation of state is below -1.  

A model with $w < -1$ requires negative kinetic energy, which in turn would require non-standard physics.  `Phantom' dark energy models with $w < -1$ have been considered via the introduction of kinetic energy that scales as $-\frac{1}{2}\dot{\phi}^2$ rather than $\frac{1}{2}\dot{\phi}^2$  \citep{Caldwell2002}, although a physical mechanism is not known.  Even this could not produce an equation of state with $w_0 = -0.831, w_a = -0.73$, since $w$ would need to cross -1, so that the kinetic energy is positive at low redshift but negative at high redshift.  It is unclear how to generate this sort of model.  However, the observational constraints on dark energy only exist in the low-redshift, $w > -1$ regime.  The $w < -1$ behavior which requires such exotic physics only appears in the extrapolation of the linear $w_ow_a$ model and does not occur in a thawing quintessence model.

A reasonable solution therefore is to use a more physically motivated model rather than a linear approximation.  The full behavior of a thawing quintessence model is computationally complex to solve.  Because the equations must be solved each time a new set of parameters is sampled, it is not practical to find an exact solution as part of an MCMC method.  

\begin{figure}[h]
    \centering
    \includegraphics[width=0.4\textwidth]{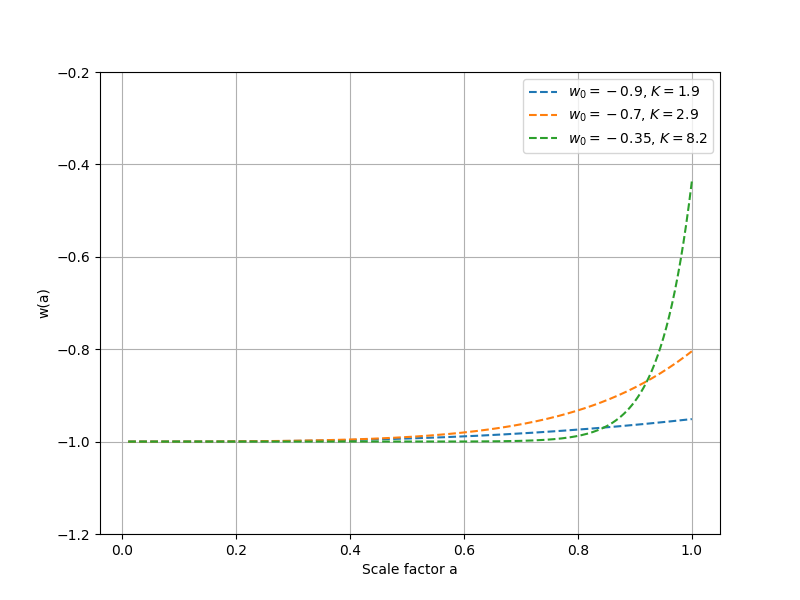}
    \caption{The equation of state as a function of scale factor with a hilltop potential for \citet{Dutta2008} approximation with three choices of parameters: $(w_0, K) \in \left\{(0.9, 1.9).(0.7,2.9),(0.35,8.2)\right\}$.  $w_0$ corresponds to the equation of state at $z = 0$ and $a = 1$, while $K$ determines whether the increase in $w$ starts earlier and is more gradual (smaller $K$) or starts later and is more rapid (larger $K$).}
    \label{fig:quintapprox} 
\end{figure}

A variety of possible approximations to quintessence models have been explored, summarized in \citet{Shlivko2025}.  The current release version of \texttt{camb} includes a $w_0w_a$ cosmology but does not include these other approximations.  Here, the \citet{Dutta2008} approximation (see also \citealt{Chiba2013}),
\begin{widetext}
\begin{equation}
    w(a)=-1+(1+w_0) a^{3(K-1)} \left[\frac{(K-F(a))(F(a)+1)^K+(K+F(a))(F(a)-1)^K}
    {(K-\Omega_{\phi 0}^{-1/2})(\Omega_{\phi 0}^{-1/2}+1)^K
    +(K+\Omega_{\phi 0}^{-1/2})(\Omega_{\phi 0}^{-1/2}-1)^K}
    \right]^2,
\end{equation}
\end{widetext}
where $w_0$ is the present value of the dark energy equation of state,
\begin{equation}
    K = \sqrt{1-\frac{4M_{\rm pl}^2 V_{,\phi \phi} (\phi_i)}
{3V(\phi_i)}},
\end{equation}
and
\begin{equation}
    F(a) = \sqrt{1+(\Omega_{\phi 0}^{-1}-1)a^{-3}},
\end{equation}
is chosen and implemented in order to better consider whether thawing quintessence can produce joint inference.  This is a two-parameter fit, with $w_0$ again the $z=0$ dark energy equation of state and $K$ controlling the epoch at which the field 'thaws' (Fig. \ref{fig:quintapprox}).  

\begin{figure}[h]
    \centering
    \includegraphics[width=0.4\textwidth]{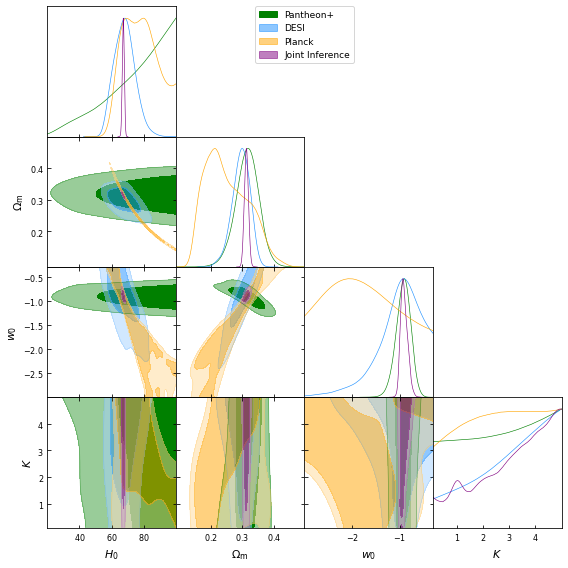}
    \caption{Each individual inference as well as the joint inference for a cosmology using the \citet{Dutta2008} approximation for the evolution of the dark energy equation of state of a thawing quintessence model.  The best-fit results include $\Lambda$CDM, for which $w_0 = -1$.}
    \label{fig:quintresults} 
\end{figure}

The best-fit parameters are shown in Fig. \ref{fig:quintresults}.  The best-fit results include $\Lambda$CDM, for which $w_0 = -1$ and $K$ can take on any value.  This more physically motivated model produces a result more consistent with $\Lambda$CDM than a linear model introducing an additional, entirely free parameter.  

An additional branch of thawing quintessence models at large $K$ is equally consistent with observations.  However, this corresponds to a model with $w_\Lambda \approx -1$ until very late times, after which there is a sharp rise in $w$ at low redshift.  Because there are essentially no constraints on the equation of state at very low redshift (e.g., in the regime where bulk flow corrections are too large), this sort of model is difficult to rule out, but is similarly not favored over $\Lambda$CDM. 

Several other approximations for varying dark energy models were also implemented (Shlivko et al. in prep.) and will be released to the broader community in a followup paper.

\section{Discussion}
\label{sec:discussion}

Current techniques to constrain cosmological parameters using the joint inference from multiple types of observations require the strong assumption that all of the measurements are mutually consistent.  However, recent observations are instead mutually inconsistent under $\Lambda$CDM, leading to a variety of invalid inferences when performing a joint analysis.  

The best solution depends upon the origin of the inconsistency between the various observations.  Possible explanations can be divided into three categories, each of which presents unique data analysis challenges.  

\begin{itemize}
    \item {If well-measured constraints are truly inconsistent under $\Lambda$CDM, a new cosmological model would be required to bring measurements back into concordance.  This would be the most exciting explanation, as it would likely lead to the discovery of additional properties of dark energy or dark matter.  However, given the relatively small number of measurements, occurring at very different epochs, a large family of models will all be able to produce concordance.  
    
    For example, the dark energy equation of state is primarily constrained from two sources, local Hubble diagram measurements and BAO observations at $z \sim 0.5-1$, since dark energy becomes negligible at higher redshifts. Thus, it is not too surprising that both a linear $w_0w_a$ model and a two-parameter thawing quintessence model can fit both measurements.  The choice of model then cannot be purely data-driven, but instead requires a subjective determination about which classes of models are most natural.}

    \item {Alternatively, the inconsistency might be due to underestimated uncertainties.  In particular, it is common to express cosmological measurements in terms of a covariance matrix, which cannot properly account for systematic uncertainties.  Thus, it might still be possible that all observations are consistent with a single set of $\Lambda$CDM parameters.  If so, the underestimated uncertainties in cosmological constraints will also lead to underestimated uncertainties in the resulting $\Lambda$CDM parameters, as shown in \S~\ref{sec:systematic}.  
    
    A resampling approach is proposed here in an attempt to fix this problem, with the idea that a non-parametric method will be independent of the reported uncertainties.  Although the reliance on reported uncertainties is removed, it is replaced by the strong assumptions that the chosen model is correct and that measured parameters are all drawn from a common distribution about the true parameters.  In particular, because resampling assumes a given model as the ground truth, it would be impossible to falsify, e.g., $\Lambda$CDM via a joint inference based on resampling.}

    \item{Finally, it could be that a single incorrect measurement is primarily responsible.  In that case, incorrect measurements should be disregarded and only the remaining measurements should be used.  Although the principle is simple, it is difficult to implement.  The Hubble tension does not arise from a single observational result, but rather from a disagreement between several internally consistent early Universe measurements and multiple other internally consistent local Universe measurements.  To resolve it, all of the measurements using a technique would need to be discarded.  Lacking clear evidence of flaw in one of the measurements, the choice of which to discard is a subjective decision, and the resulting $\Lambda$CDM parameters (or even whether the model can be falsified) will depend strongly upon that choice.
    
    Further, the full $\Lambda$CDM model can only be constrained from a combination of measurements at different cosmological epochs.  By removing some of the constraints, the resulting inference depends strongly on the priors and bounds on model parameters (such as in Fig. \ref{fig:bounds}, Table \ref{tab:bounds}).}
\end{itemize}

The goal of using cosmological measurements to test $\Lambda$CDM should be an objective determination of whether they are consistent with a single set of $\Lambda$CDM parameters or whether the model can be falsified.  However, the current situation instead requires multiple subjective decisions in order to produce a joint inference.  First, one must decide whether to interpret inconsistent measurements as due to features of the measurements or true inconsistency with the model.  As long as the former is a plausible interpretation, the model can never be falsified.  If the latter is chosen, additional assumptions are required to bring seemingly inconsistent measurements into agreement.  The eventual conclusion is then strongly dependent upon all of these assumptions.

\subsection{The Flying Unicorn Problem}

Surprisingly, even the choice of model itself can present a complicated falsifiability problem.  Because the MCMC walkers explore the full parameter space available to them, the choice of bounds even far from the minimum residual can affect the results.  Depending upon which combinations of parameters are available to explore, some measurements may effectively be overweighted or underweighted, altering the resulting best-fit parameters.  The effect is particularly strong when there are otherwise insufficient constraints, such as the sensitivity of the best-fit $w_0w_a$ parameters for SNe constraints alone as in Fig. \ref{fig:bounds}.

The same issue occurs when additional parameters, even purely nuisance parameters, are added to the model.  For example, suppose that in addition to the standard $\Lambda$CDM components of $w = -1$ dark energy and $w = 0$ matter, one decides that perhaps a third component, flying unicorns (in order to have $w_{\unicorn} > 0$, naturally unicorns should be assumed to fly at relativistic speeds), should be added.  The same Bayesian joint inference can be performed, and it turns out that the best-fit value of $\Omega_{\unicorn}$ is consistent with zero.  Thus, there is no evidence for flying unicorns in our observable Universe.  

However, as with changing the bounds, the best-fit $\Omega_m$ and $\Omega_\Lambda$ also change, because additional parameter space is explored by the walkers.  This presents a significant philosophical problem when determining which values of $\Omega_m$ and $\Omega_\Lambda$ to use.  One might imagine a debate between two cosmologists, one favoring the standard $\Lambda$CDM model and the other favoring $\unicorn\Lambda$CDM.  They disagree on whether flying unicorns are plausible.  However, they both agree that $\Omega_{\unicorn} = 0$ and on every other observable.  The difference is simply that the $\unicorn\Lambda$CDM proponent believes that flying unicorns could exist, and simply happen not to be in our observable Universe, whereas the $\Lambda$CDM proponent believes they cannot exist.  And that, unfortunately, is insufficient basis to falsify $\unicorn\Lambda$CDM.  

To use a more concrete example, it is not only the $w_0w_a$ model which is sensitive to the chosen bounds.  When fitting $\Lambda$CDM parameters to SNe data, the result is sensitive to bounds on $\Omega_m$.  The default \texttt{cobaya} bound is $\Omega_m > 0.1$, and the physical model breaks down for $\Omega_m < \Omega_b$.  However, if the bounds are extended to allow walkers to explore $\Omega_m < 0$, it lowers the resulting $\Omega_m$.  The best-fit answer is still $\Omega_m \sim 0.2$, a positive value.  But, like with $\unicorn\Lambda$CDM, this means that there will be a disagreement between cosmologists who believe $\Omega_m$ cannot be negative and those who believe it could be, and just happens not to be, and one cannot use cosmological constraints to falsify the latter interpretation.  For such unicorn problems, one must again make a subjective decision on which models are reasonable, or natural, or simpler, or physically motivated rather than an objective determination that ideas have been falsified.  

\subsection{Consequences for $\Lambda$CDM}

If a subjective determination is ultimately required, then perhaps it is worth re-examining how strongly current evidence argues for a non-$\Lambda$CDM cosmology.  As outlined here, there are three possible reasons that current observations might be inconsistent with $\Lambda$CDM.  

If the disagreement is due to an erroneous measurement that must be discarded, the remaining constraints can be used to probe $\Lambda$CDM.  As shown, this leads to a situation in which the result is often strongly dependent upon the priors and the bounds set on MCMC walkers.  In particular, the result that $w_0 \neq -1, w_a \neq 0$ for a $w_0w_a$ model is strongest when allowing $w_0$ and $w_a$ to reach large negative values.  However, these are the least interpretable models, since for $w < -1$ the null energy condition is violated.  As the search space is more closely restricted to models satsifying the null energy condition, the result is increasingly consistent with $\Lambda$CDM.  

Further, even if models violating the null energy condition are allowed, the resulting fits only produce $w < -1$ in a regime which is unconstrained by observations.  At every epoch where it can be tested, $w \geq -1$.  Thus, $w < -1$ cosmological models are a version of the unicorn problem described above.

If the disagreement is instead due to underestimated or systematic errors not included in the reported likelihoods, the most promising approach requires the assumption that an imposed model is correct.  Thus, it would not be possible to falsify $\Lambda$CDM or any other model without first producing robust probability distributions for each cosmological constraint.

Finally, one must consider the possibility that a new model is truly required.  Here, perhaps the most natural current candidate, a thawing quintessence model, is implemented in \texttt{cobaya} and considered.  The best-fit quintessence parameters have a large $K$, indicating that the deviation between the model and $\Lambda$CDM only occurs at late enough times that there are no direct constraints.  In addition, the present value $w_0$ is consistent with -1.  Thus, by introducing a physically-motivated alternative model rather than one that violates the null energy condition, the result becomes more consistent with $\Lambda$CDM.  

Ultimately, since a subjective determination is unfortunately required, the situation is open to interpretation.  In each case, attempts to improve the analysis appear to produce results that are increasingly consistent with $\Lambda$CDM.  A reasonable interpretation is that $\Lambda$CDM should not yet be rejected on the basis of current measurements.  

\subsection{Guidelines for Cosmological Inference}

As with any scientific model, it is likely that at some point, experimental or observational evidence will convincingly falsify $\Lambda$CDM and the model will be improved.  When that happens, it is essential to use mathematically sound analysis techniques that produce a robust answer.  This is by no means straightforward for measurements using a variety of techniques and relying on a variety of strong assumptions to measure different observables at a variety of cosmological epochs.  Although it is difficult to rigorously validate a joint inference technique for the reasons described in this work, many invalid techniques share two common properties which can be examined.

First, any robust answer must be only weakly sensitive to choices of priors and bounds.  If a change in bounds far from the best-fit parameters is the difference between a result that is consistent or inconsistent with a model, then the technique cannot be used to falsify that model.

Second, the model itself is another sort of prior and should be considered in a similar manner.  If the resulting fit is strongly sensitive to introduction of a `unicorn' into the model, it will not be possible to precisely determine cosmological parameters.  

There are several paths to producing a joint inference satisfying both conditions.  Certainly the preferred path would be improving our understanding of the uncertainties associated with each cosmological measurement.  After all, no scientific measurement is meaningful due to the central value alone.  However, historically it has proven difficult to properly bound uncertainties in cosmological measurements, particularly those using the distance ladder.  

An alternative might be to find additional measurement techniques so that multiple observations are available at the same cosmological epoch, ideally also measuring the same set of parameters.  In that case, even if the uncertainties are not individually well constrained, resampling would truly allow a non-parametric estimate of the uncertainty distribution, since there would no longer be the need to assume a particular cosmological model in order to connect constraints at different epochs.

The authors would like to thank Jens Hjorth, David Shlivko, and Paul Steinhardt for helpful comments.  CS, PP, and RW were supported by research grants (VIL16599, VIL54489) from VILLUM FONDEN. PP was supported by UT's Kushner Experiential Learning Fellowship.

\bibliographystyle{aasjournal}
\bibliography{refs.bib} 

\begin{thebibliography}{}
\expandafter\ifx\csname natexlab\endcsname\relax\def\natexlab#1{#1}\fi
\providecommand{\url}[1]{\href{#1}{#1}}
\providecommand{\dodoi}[1]{doi:~\href{http://doi.org/#1}{\nolinkurl{#1}}}
\providecommand{\doeprint}[1]{\href{http://ascl.net/#1}{\nolinkurl{http://ascl.net/#1}}}
\providecommand{\doarXiv}[1]{\href{https://arxiv.org/abs/#1}{\nolinkurl{https://arxiv.org/abs/#1}}}

\bibitem[{{Abbott} {et~al.}(2024){Abbott}, {Acevedo}, {Aguena}, {Alarcon}, {Allam}, {Alves}, {Amon}, {Andrade-Oliveira}, {Annis}, {Armstrong}, {Asorey}, {Avila}, {Bacon}, {Bassett}, {Bechtol}, {Bernardinelli}, {Bernstein}, {Bertin}, {Blazek}, {Bocquet}, {Brooks}, {Brout}, {Buckley-Geer}, {Burke}, {Camacho}, {Camilleri}, {Campos}, {Carnero Rosell}, {Carollo}, {Carr}, {Carretero}, {Castander}, {Cawthon}, {Chang}, {Chen}, {Choi}, {Conselice}, {Costanzi}, {da Costa}, {Crocce}, {Davis}, {DePoy}, {Desai}, {Diehl}, {Dixon}, {Dodelson}, {Doel}, {Doux}, {Drlica-Wagner}, {Elvin-Poole}, {Everett}, {Ferrero}, {Fert{\'e}}, {Flaugher}, {Foley}, {Fosalba}, {Friedel}, {Frieman}, {Frohmaier}, {Galbany}, {Garc{\'\i}a-Bellido}, {Gatti}, {Gaztanaga}, {Giannini}, {Glazebrook}, {Graur}, {Gruen}, {Gruendl}, {Gutierrez}, {Hartley}, {Herner}, {Hinton}, {Hollowood}, {Honscheid}, {Huterer}, {Jain}, {James}, {Jeffrey}, {Kasai}, {Kelsey}, {Kent}, {Kessler}, {Kim}, {Kirshner}, {Kovacs}, {Kuehn}, {Lahav}, {Lee}, {Lee}, {Lewis}, {Li},
  {Lidman}, {Lin}, {Malik}, {Marshall}, {Martini}, {Mena-Fern{\'a}ndez}, {Menanteau}, {Miquel}, {Mohr}, {Mould}, {Muir}, {M{\"o}ller}, {Neilsen}, {Nichol}, {Nugent}, {Ogando}, {Palmese}, {Pan}, {Paterno}, {Percival}, {Pereira}, {Pieres}, {Plazas Malag{\'o}n}, {Popovic}, {Porredon}, {Prat}, {Qu}, {Raveri}, {Rodr{\'\i}guez-Monroy}, {Romer}, {Roodman}, {Rose}, {Sako}, {Sanchez}, {Sanchez Cid}, {Schubnell}, {Scolnic}, {Sevilla-Noarbe}, {Shah}, {Smith}, {Smith}, {Soares-Santos}, {Suchyta}, {Sullivan}, {Suntzeff}, {Swanson}, {S{\'a}nchez}, {Tarle}, {Taylor}, {Thomas}, {To}, {Toy}, {Troxel}, {Tucker}, {Tucker}, {Uddin}, {Vincenzi}, {Walker}, {Weaverdyck}, {Wechsler}, {Weller}, {Wester}, {Wiseman}, {Yamamoto}, {Yuan}, {Zhang}, {Zhang}, \& {DES Collaboration}}]{Abbott2024}
{Abbott}, T.~M.~C., {Acevedo}, M., {Aguena}, M., {et~al.} 2024, \apjl, 973, L14, \dodoi{10.3847/2041-8213/ad6f9f}

\bibitem[{{Blas} {et~al.}(2011){Blas}, {Lesgourgues}, \& {Tram}}]{Blas2011}
{Blas}, D., {Lesgourgues}, J., \& {Tram}, T. 2011, \jcap, 2011, 034, \dodoi{10.1088/1475-7516/2011/07/034}

\bibitem[{{Bolte} \& {Hogan}(1995)}]{Bolte1995}
{Bolte}, M., \& {Hogan}, C.~J. 1995, \nat, 376, 399, \dodoi{10.1038/376399a0}

\bibitem[{{Bond} {et~al.}(2013){Bond}, {Nelan}, {VandenBerg}, {Schaefer}, \& {Harmer}}]{Bond2013}
{Bond}, H.~E., {Nelan}, E.~P., {VandenBerg}, D.~A., {Schaefer}, G.~H., \& {Harmer}, D. 2013, \apjl, 765, L12, \dodoi{10.1088/2041-8205/765/1/L12}

\bibitem[{{Caldwell}(2002)}]{Caldwell2002}
{Caldwell}, R.~R. 2002, Physics Letters B, 545, 23, \dodoi{10.1016/S0370-2693(02)02589-3}

\bibitem[{{Chevallier} \& {Polarski}(2001)}]{Chevallier2001}
{Chevallier}, M., \& {Polarski}, D. 2001, International Journal of Modern Physics D, 10, 213, \dodoi{10.1142/S0218271801000822}

\bibitem[{{Chiba} {et~al.}(2013){Chiba}, {De Felice}, \& {Tsujikawa}}]{Chiba2013}
{Chiba}, T., {De Felice}, A., \& {Tsujikawa}, S. 2013, \prd, 87, 083505, \dodoi{10.1103/PhysRevD.87.083505}

\bibitem[{{de Putter} \& {Linder}(2008)}]{dePutter2008}
{de Putter}, R., \& {Linder}, E.~V. 2008, \jcap, 2008, 042, \dodoi{10.1088/1475-7516/2008/10/042}

\bibitem[{{DESI Collaboration} {et~al.}(2024){DESI Collaboration}, {Adame}, {Aguilar}, {Ahlen}, {Alam}, {Alexander}, {Alvarez}, {Alves}, {Anand}, {Andrade}, {Armengaud}, {Avila}, {Aviles}, {Awan}, {Bahr-Kalus}, {Bailey}, {Baltay}, {Bault}, {Behera}, {BenZvi}, {Bera}, {Beutler}, {Bianchi}, {Blake}, {Blum}, {Brieden}, {Brodzeller}, {Brooks}, {Buckley-Geer}, {Burtin}, {Calderon}, {Canning}, {Carnero Rosell}, {Cereskaite}, {Cervantes-Cota}, {Chabanier}, {Chaussidon}, {Chaves-Montero}, {Chen}, {Chen}, {Claybaugh}, {Cole}, {Cuceu}, {Davis}, {Dawson}, {de la Macorra}, {de Mattia}, {Deiosso}, {Dey}, {Dey}, {Ding}, {Doel}, {Edelstein}, {Eftekharzadeh}, {Eisenstein}, {Elliott}, {Fagrelius}, {Fanning}, {Ferraro}, {Ereza}, {Findlay}, {Flaugher}, {Font-Ribera}, {Forero-S{\'a}nchez}, {Forero-Romero}, {Frenk}, {Garcia-Quintero}, {Gazta{\~n}aga}, {Gil-Mar{\'\i}n}, {Gontcho}, {Gonzalez-Morales}, {Gonzalez-Perez}, {Gordon}, {Green}, {Gruen}, {Gsponer}, {Gutierrez}, {Guy}, {Hadzhiyska}, {Hahn}, {Hanif}, {Herrera-Alcantar},
  {Honscheid}, {Howlett}, {Huterer}, {Ir{\v{s}}i{\v{c}}}, {Ishak}, {Juneau}, {Kara{\c{c}}ayl{\i}}, {Kehoe}, {Kent}, {Kirkby}, {Kremin}, {Krolewski}, {Lai}, {Lan}, {Landriau}, {Lang}, {Lasker}, {Le Goff}, {Le Guillou}, {Leauthaud}, {Levi}, {Li}, {Linder}, {Lodha}, {Magneville}, {Manera}, {Margala}, {Martini}, {Maus}, {McDonald}, {Medina-Varela}, {Meisner}, {Mena-Fern{\'a}ndez}, {Miquel}, {Moon}, {Moore}, {Moustakas}, {Mudur}, {Mueller}, {Mu{\~n}oz-Guti{\'e}rrez}, {Myers}, {Nadathur}, {Napolitano}, {Neveux}, {Newman}, {Nguyen}, {Nie}, {Niz}, {Noriega}, {Padmanabhan}, {Paillas}, {Palanque-Delabrouille}, {Pan}, {Penmetsa}, {Percival}, {Pieri}, {Pinon}, {Poppett}, {Porredon}, {Prada}, {P{\'e}rez-Fern{\'a}ndez}, {P{\'e}rez-R{\`a}fols}, {Rabinowitz}, {Raichoor}, {Ram{\'\i}rez-P{\'e}rez}, {Ramirez-Solano}, {Ravoux}, {Rashkovetskyi}, {Rezaie}, {Rich}, {Rocher}, {Rockosi}, {Roe}, {Rosado-Marin}, {Ross}, {Rossi}, {Ruggeri}, {Ruhlmann-Kleider}, {Samushia}, {Sanchez}, {Saulder}, {Schlafly}, {Schlegel}, {Schubnell}, {Seo},
  {Shafieloo}, {Sharples}, {Silber}, {Slosar}, {Smith}, {Sprayberry}, {Tan}, {Tarl{\'e}}, {Taylor}, {Trusov}, {Ure{\~n}a-L{\'o}pez}, {Vaisakh}, {Valcin}, {Valdes}, {Vargas-Maga{\~n}a}, {Verde}, {Walther}, {Wang}, {Wang}, {Weaver}, {Weaverdyck}, {Wechsler}, {Weinberg}, {White}, {Yu}, {Yu}, {Yuan}, {Y{\`e}che}, {Zaborowski}, {Zarrouk}, {Zhang}, {Zhao}, {Zhao}, {Zhou}, {Zhuang}, \& {Zou}}]{DESIY1}
{DESI Collaboration}, {Adame}, A.~G., {Aguilar}, J., {et~al.} 2024, arXiv e-prints, arXiv:2404.03002, \dodoi{10.48550/arXiv.2404.03002}

\bibitem[{{Dutta} \& {Scherrer}(2008)}]{Dutta2008}
{Dutta}, S., \& {Scherrer}, R.~J. 2008, \prd, 78, 123525, \dodoi{10.1103/PhysRevD.78.123525}

\bibitem[{{Efstathiou} \& {Bond}(1999)}]{Efstathiou1999}
{Efstathiou}, G., \& {Bond}, J.~R. 1999, \mnras, 304, 75, \dodoi{10.1046/j.1365-8711.1999.02274.x}

\bibitem[{{Foreman-Mackey} {et~al.}(2013){Foreman-Mackey}, {Hogg}, {Lang}, \& {Goodman}}]{ForemanMackey2013}
{Foreman-Mackey}, D., {Hogg}, D.~W., {Lang}, D., \& {Goodman}, J. 2013, \pasp, 125, 306, \dodoi{10.1086/670067}

\bibitem[{{Lewis} \& {Bridle}(2002)}]{Lewis2002}
{Lewis}, A., \& {Bridle}, S. 2002, \prd, 66, 103511, \dodoi{10.1103/PhysRevD.66.103511}

\bibitem[{Lewis {et~al.}(2000)Lewis, Challinor, \& Lasenby}]{Lewis1999}
Lewis, A., Challinor, A., \& Lasenby, A. 2000, \apj, 538, 473, \dodoi{10.1086/309179}

\bibitem[{{Li} {et~al.}(2005){Li}, {Feng}, \& {Zhang}}]{Li2005}
{Li}, M., {Feng}, B., \& {Zhang}, X. 2005, \jcap, 2005, 002, \dodoi{10.1088/1475-7516/2005/12/002}

\bibitem[{{Linder}(2003)}]{Linder2003}
{Linder}, E.~V. 2003, \prl, 90, 091301, \dodoi{10.1103/PhysRevLett.90.091301}

\bibitem[{{Metropolis} {et~al.}(1953){Metropolis}, {Rosenbluth}, {Rosenbluth}, {Teller}, \& {Teller}}]{Metropolis1953}
{Metropolis}, N., {Rosenbluth}, A.~W., {Rosenbluth}, M.~N., {Teller}, A.~H., \& {Teller}, E. 1953, \jcp, 21, 1087, \dodoi{10.1063/1.1699114}

\bibitem[{{Ostriker} \& {Steinhardt}(1995)}]{Ostriker1995}
{Ostriker}, J.~P., \& {Steinhardt}, P.~J. 1995, \nat, 377, 600, \dodoi{10.1038/377600a0}

\bibitem[{{Perlmutter} {et~al.}(1999){Perlmutter}, {Aldering}, {Goldhaber}, {Knop}, {Nugent}, {Castro}, {Deustua}, {Fabbro}, {Goobar}, {Groom}, {Hook}, {Kim}, {Kim}, {Lee}, {Nunes}, {Pain}, {Pennypacker}, {Quimby}, {Lidman}, {Ellis}, {Irwin}, {McMahon}, {Ruiz-Lapuente}, {Walton}, {Schaefer}, {Boyle}, {Filippenko}, {Matheson}, {Fruchter}, {Panagia}, {Newberg}, {Couch}, \& {Project}}]{Perlmutter1999}
{Perlmutter}, S., {Aldering}, G., {Goldhaber}, G., {et~al.} 1999, \apj, 517, 565, \dodoi{10.1086/307221}

\bibitem[{{Planck Collaboration} {et~al.}(2020){Planck Collaboration}, {Aghanim}, {Akrami}, {Ashdown}, {Aumont}, {Baccigalupi}, {Ballardini}, {Banday}, {Barreiro}, {Bartolo}, {Basak}, {Battye}, {Benabed}, {Bernard}, {Bersanelli}, {Bielewicz}, {Bock}, {Bond}, {Borrill}, {Bouchet}, {Boulanger}, {Bucher}, {Burigana}, {Butler}, {Calabrese}, {Cardoso}, {Carron}, {Challinor}, {Chiang}, {Chluba}, {Colombo}, {Combet}, {Contreras}, {Crill}, {Cuttaia}, {de Bernardis}, {de Zotti}, {Delabrouille}, {Delouis}, {Di Valentino}, {Diego}, {Dor{\'e}}, {Douspis}, {Ducout}, {Dupac}, {Dusini}, {Efstathiou}, {Elsner}, {En{\ss}lin}, {Eriksen}, {Fantaye}, {Farhang}, {Fergusson}, {Fernandez-Cobos}, {Finelli}, {Forastieri}, {Frailis}, {Fraisse}, {Franceschi}, {Frolov}, {Galeotta}, {Galli}, {Ganga}, {G{\'e}nova-Santos}, {Gerbino}, {Ghosh}, {Gonz{\'a}lez-Nuevo}, {G{\'o}rski}, {Gratton}, {Gruppuso}, {Gudmundsson}, {Hamann}, {Handley}, {Hansen}, {Herranz}, {Hildebrandt}, {Hivon}, {Huang}, {Jaffe}, {Jones}, {Karakci}, {Keih{\"a}nen},
  {Keskitalo}, {Kiiveri}, {Kim}, {Kisner}, {Knox}, {Krachmalnicoff}, {Kunz}, {Kurki-Suonio}, {Lagache}, {Lamarre}, {Lasenby}, {Lattanzi}, {Lawrence}, {Le Jeune}, {Lemos}, {Lesgourgues}, {Levrier}, {Lewis}, {Liguori}, {Lilje}, {Lilley}, {Lindholm}, {L{\'o}pez-Caniego}, {Lubin}, {Ma}, {Mac{\'\i}as-P{\'e}rez}, {Maggio}, {Maino}, {Mandolesi}, {Mangilli}, {Marcos-Caballero}, {Maris}, {Martin}, {Martinelli}, {Mart{\'\i}nez-Gonz{\'a}lez}, {Matarrese}, {Mauri}, {McEwen}, {Meinhold}, {Melchiorri}, {Mennella}, {Migliaccio}, {Millea}, {Mitra}, {Miville-Desch{\^e}nes}, {Molinari}, {Montier}, {Morgante}, {Moss}, {Natoli}, {N{\o}rgaard-Nielsen}, {Pagano}, {Paoletti}, {Partridge}, {Patanchon}, {Peiris}, {Perrotta}, {Pettorino}, {Piacentini}, {Polastri}, {Polenta}, {Puget}, {Rachen}, {Reinecke}, {Remazeilles}, {Renzi}, {Rocha}, {Rosset}, {Roudier}, {Rubi{\~n}o-Mart{\'\i}n}, {Ruiz-Granados}, {Salvati}, {Sandri}, {Savelainen}, {Scott}, {Shellard}, {Sirignano}, {Sirri}, {Spencer}, {Sunyaev}, {Suur-Uski}, {Tauber}, {Tavagnacco},
  {Tenti}, {Toffolatti}, {Tomasi}, {Trombetti}, {Valenziano}, {Valiviita}, {Van Tent}, {Vibert}, {Vielva}, {Villa}, {Vittorio}, {Wandelt}, {Wehus}, {White}, {White}, {Zacchei}, \& {Zonca}}]{Planck2020}
{Planck Collaboration}, {Aghanim}, N., {Akrami}, Y., {et~al.} 2020, \aap, 641, A6, \dodoi{10.1051/0004-6361/201833910}

\bibitem[{{Riess} {et~al.}(1998){Riess}, {Filippenko}, {Challis}, {Clocchiatti}, {Diercks}, {Garnavich}, {Gilliland}, {Hogan}, {Jha}, {Kirshner}, {Leibundgut}, {Phillips}, {Reiss}, {Schmidt}, {Schommer}, {Smith}, {Spyromilio}, {Stubbs}, {Suntzeff}, \& {Tonry}}]{Riess1998}
{Riess}, A.~G., {Filippenko}, A.~V., {Challis}, P., {et~al.} 1998, \aj, 116, 1009, \dodoi{10.1086/300499}

\bibitem[{{Riess} {et~al.}(2022){Riess}, {Yuan}, {Macri}, {Scolnic}, {Brout}, {Casertano}, {Jones}, {Murakami}, {Anand}, {Breuval}, {Brink}, {Filippenko}, {Hoffmann}, {Jha}, {D'arcy Kenworthy}, {Mackenty}, {Stahl}, \& {Zheng}}]{Riess2022}
{Riess}, A.~G., {Yuan}, W., {Macri}, L.~M., {et~al.} 2022, \apjl, 934, L7, \dodoi{10.3847/2041-8213/ac5c5b}

\bibitem[{{Scolnic} {et~al.}(2022){Scolnic}, {Brout}, {Carr}, {Riess}, {Davis}, {Dwomoh}, {Jones}, {Ali}, {Charvu}, {Chen}, {Peterson}, {Popovic}, {Rose}, {Wood}, {Brown}, {Chambers}, {Coulter}, {Dettman}, {Dimitriadis}, {Filippenko}, {Foley}, {Jha}, {Kilpatrick}, {Kirshner}, {Pan}, {Rest}, {Rojas-Bravo}, {Siebert}, {Stahl}, \& {Zheng}}]{Scolnic2022}
{Scolnic}, D., {Brout}, D., {Carr}, A., {et~al.} 2022, \apj, 938, 113, \dodoi{10.3847/1538-4357/ac8b7a}

\bibitem[{{Shlivko} {et~al.}(2025){Shlivko}, {Steinhardt}, \& {Steinhardt}}]{Shlivko2025}
{Shlivko}, D., {Steinhardt}, P.~J., \& {Steinhardt}, C.~L. 2025, arXiv e-prints, arXiv:2504.02028.
\newblock \doarXiv{2504.02028}

\bibitem[{{Torrado} \& {Lewis}(2021)}]{Torrado2021}
{Torrado}, J., \& {Lewis}, A. 2021, \jcap, 2021, 057, \dodoi{10.1088/1475-7516/2021/05/057}

\bibitem[{{Tsujikawa}(2013)}]{Tsujikawa2013}
{Tsujikawa}, S. 2013, Classical and Quantum Gravity, 30, 214003, \dodoi{10.1088/0264-9381/30/21/214003}

\bibitem[{{Tully}(2023)}]{Tully2023}
{Tully}, R.~B. 2023, arXiv e-prints, arXiv:2305.11950, \dodoi{10.48550/arXiv.2305.11950}

\end{thebibliography}


\label{lastpage}
\end{document}